\documentclass[12pt]{article}
\usepackage{epsfig}
\usepackage{cite}
\usepackage{mcite}
\usepackage{url}

\usepackage{array,tabularx,epsfig,mathrsfs,graphicx,rotating}
\usepackage{ifthen}
\usepackage{fancybox}
\usepackage{amsfonts}

\usepackage[pagewise]{lineno}

\newcommand{\beq}{\begin{equation}}
\newcommand{\eeq}{\end{equation}}

\chardef\til=126

\begin{document}

\clearpage
\pagestyle{empty}
\setcounter{footnote}{0}\setcounter{page}{0}%
\thispagestyle{empty}\pagestyle{plain}\pagenumbering{arabic}%

\hfill  ANL-HEP-PR-11-63 

\hfill  December 3, 2013

\vspace{2.0cm}

\begin{center}

{\Large\bf
A non-parametric peak finder algorithm and its application in 
searches for new physics 
\\[-1cm] }

\vspace{2.5cm}

{\large S.V.~Chekanov and M.~Erickson\footnote[1]{Also affiliated with the  Physics Department, The College of New Jersey, New Jersey, USA.} 

\vspace{0.5cm}
\itemsep=-1mm
\normalsize
\small
HEP Division, Argonne National Laboratory,
9700 S.Cass Avenue, \\ 
Argonne, IL 60439
USA
}

\normalsize
\vspace{1.0cm}


\vspace{0.5cm}
\begin{abstract}
We have developed an algorithm for non-parametric 
fitting and extraction of statistically significant peaks in the presence
of statistical and systematic uncertainties.
Applications of this algorithm for analysis of 
high-energy collision data are discussed. In  particular, we illustrate 
how to use this algorithm  in  general searches
for new physics in invariant-mass spectra using 
$pp$ Monte Carlo simulations.
\end{abstract}

\end{center}

\newpage
\setcounter{page}{1}


\section{Introduction}

Searches for peaks in particle spectra is a task 
which is becoming increasingly popular at the Large-Hadron collider that focuses on new physics beyond TeV-scale.
Bump searches can be performed either in single-particle (such as $p_T$ distributions) or invariant-mass spectra. 
For instance, searches for new particles decaying into a two-body final state (jet-jet, gamma-gamma, etc.) and multi-body
decays are typically done by examining invariant masses of final-state objects (jets, leptons, missing transverse
momenta, etc.). For example, assuming  seven identified particles (jets,  photons,
electrons, muons, taus, $Z$-bosons, missing $p_T$),
a search can be made for parent particles decaying into 2, 3, or 4 daughter
particles. This leads to 322 unique daughter groups. 
Thus, the task of analyzing such invariant-mass combinations 
becomes rather tedious and difficult to handle. 
Considering a ``blind'' analysis techniques for scanning many channels \cite{Choudalakis:2008pr}, 
any cut variation  increases the number of channels that need
to be investigated.
Finally, similar challenges exist  for automatic searches for new hadronic resonances combining tracks \cite{cppsearch}. 
   
The  task of finding bumps is
ultimately related to the task of determining  a correct background shape using theoretical or known 
cross sections.  However, a theory   
can be rather uncertain in the regions of interest, difficult to use for background simulation 
or entirely nonexistent.  
Even for a simple jet-jet invariant mass, finding an analytical background function that
fits the QCD-driven background spanning many orders in magnitude and which can be used to extract 
possible excess of events due to new physics
requires a careful examination.  Attempts to fit two-jet and three-jet 
invariant masses  have been discussed in CMS \cite{PhysRevLett.105.211801,Chatrchyan:2011cj} 
and ATLAS \cite{Aad:2011aj} papers; while both experiments have
reached the necessary precision for such fits using initial low-statistics data, the used analytical functions are 
rather different and have many free parameters. 
This task becomes even more difficult considering multiple
channels (invariant-mass distributions) with various cuts or detector-selection criteria (like $b$-tagging). 
Each such channel requires a careful selection of analytical functions for background fit and adjustments of 
their initial values  
for convergence of a non-linear regression while determining  an expectantly smooth background shape.  
A fully automated
approach to searches for new physics has been discussed elsewhere \cite{Aaltonen:2008vt}. 

One technically attractive approach is to find a non-parametric way to extract statistically significant peaks without {\em a prior}
assumptions on background shapes. 
Such approach is popular in many areas, from image processing to studies of financial market, where a  
typical peak-identification task is reduced to data smoothing in order to create a 
function that attempts to approximate the background. 
The smoothing can be achieved  using the moving average \cite{Mav}, Lowess \cite{lowe}, SPlines \cite{SPline} algorithms.
Statistically significant deviations from smoothed distributions can be considered as 
peaks. 
Such technique is certainly adequate for the peak extraction, but it does not 
pursue the goal of peak identification with a correct treatment of statistical (or systematic) uncertainties.
The later can be asymmetric.

The closest peak-search approach for high-energy-physics applications
has been developed for studies of $\gamma$-ray spectra where the usual features of interest are the energies and intensities of photo-peaks. 
Several techniques have been developed, such as those based on  
least squares \cite{Huang1969141}, second differences with least-squares fitting \cite{Mariscotti1967309},
Fourier transformation \cite{Blinowska1974597}, Markov chain \cite{Silagadze:1995zd}, convolution \cite{2000NIMPA.443..108M},
(just naming a few). 
While such approaches are well suited for counting-type observables, they typically focus on narrow peaks on top of small
and often flat-shaped background. 

For example, the ROOT analysis framework \cite{root} used in high-energy physics 
contains the {\sc TSpectrum} package based on a smoothing 
method developed for $\gamma$-ray spectra \cite{root_peak}. The latter typically have narrow peaks on a smooth  background. 
This algorithm is efficient in finding sharp peaks, while detection of wide peaks requires a visual examination of data 
to adjust of several free parameters of this tool.
Thus this approach is not well suited for a completely automatic peak search. In addition, systematic uncertainties
on data points are not easy to incorporate in this approach.

In high-energy collisions, a typical Standard-Model background distribution has a falling shape  
spanning many orders of magnitude in event counts. 
A typical example is jet-jet invariant masses used for new particle searches \cite{PhysRevLett.105.211801,Aad:2011aj}. For such spectra,  
the  most  interesting regions are the tails of the exponentially suppressed distributions where a new high-$p_T$ physics may show up.
This means that there should be rather  different  thresholds to statistical noise, depending on the phase-space region, and
as the result, a correct treatment of statistical and systematic uncertainties is obligatory.
Unlike the $\gamma$-ray spectra where peaks are rather common and subject of various classification techniques, 
peaks in high-energy collisions are rather rare. As a consequence,  
relatively little progress has been made to develop a non-parametric fitting technique for high-energy physics applications where an observation of peaks 
is typically a subject for searches for new physics 
rather than for peak-classification purposes.

The above discussion leads to the need for a non-parametric way of background estimation together with
the peak extraction mechanism which can be suited for high-energy collision distributions, such as invariant masses.
The algorithm should be able to  take into account the discrete nature of input distributions  
with their uncertainties. The proposed algorithm is less ambiguous compared to the smoothing methods (such as
that used in ROOT \cite{root,root_peak}), since it uses only one free parameter. In addition, it can take into account systematic uncertainties on data points (that can be asymmetric),
and thus can estimate statistical significance of possible peaks in the presence of systematic uncertainties.

\section{Non-parametric peak finder algorithm}

Due to the reasons discussed above, the
program called Non-Parametric Peak Finder (NPFinder) was developed using a numerical, iterative
approach to detect statistically significant peaks in event-counting distributions. In short, NPFinder iterates through bins 
of input histograms and, using only one sensitivity parameter,
determines the location and statistical significance of possible peaks. Unlike the known smoothing algorithms,
the main focus of this method is not how to smooth data and then extract  peaks, but rather
how to extract peaks  by comparing neighboring points and then calling what is left over the "background".
Below we discuss the major elements of this algorithm and then we illustrate and discuss its limitations and possible
improvements.

For each point $i$ in a histogram, the first-order derivative $\alpha_i$ 
is found taking into account possible (statistical or/and systematic) uncertainties. This is done by
calculating the slope between two points including their experimental uncertainties: If point $i+1$ is lower
than point $i$, the upper error is used, while if point $i+1$ is higher than point $i$, the lower error uncertainty is used. This
is done in order to be always on a conservative side while reducing statistical noise. 
Mathematically, this can be written as:

\begin{equation}
\alpha_i = \frac{y_{i+1} + \delta y_{i+1} - y_{i}}{x_{i+1}-x_{i}},
\label{q1}
\end{equation}
where the uncertainty $\delta y_{i+1}$ is taken with negative sign  for $y_{i+1}>y_{i}$, and with positive sign otherwise.
The uncertainty may not need to be symmetric; but for simplicity we assume that they are symmetric as this is 
usually the case for statistical nature of uncertainties.
The derivatives are averaged calculating a running average for any given position $N$:

\begin{equation}
\bar{\alpha}_N = \frac{1}{N}\sum^{N}_{i=0} \alpha_i.  
\label{q2}
\end{equation}
The algorithm triggers the beginning  of a peak if the local derivatives satisfy:

\begin{equation}
\begin{array}{l}
\delta\alpha_{N+1} = \alpha_{N+1}-\bar{\alpha}_N >  \Delta , \\ 
\delta\alpha_{N+2} = \alpha_{N+2}-\bar{\alpha}_N >  \Delta , 
\end{array}
\label{q3}
\end{equation}
where $\Delta$ is a free positive parameter that reflects (unknown) slope of the peak. This parameter should be
found empirically and we will discuss below a possible range for its value. 
When the above conditions are true, NPFinder registers a possible peak and begins classifying next
points as a part of the peak. The running average Eq.~\ref{q2} is not accumulated for the points which belong to a possible peak.
$\Delta$ is the only free parameter  which 
specifies the sensitivity to the peak finding. This parameter should decrease with 
increase of sensitivity to the peaks (and likely will increase sensitivity to statistical fluctuations).   

\begin{figure}[htp]
\begin{center}
\mbox{\epsfig{file=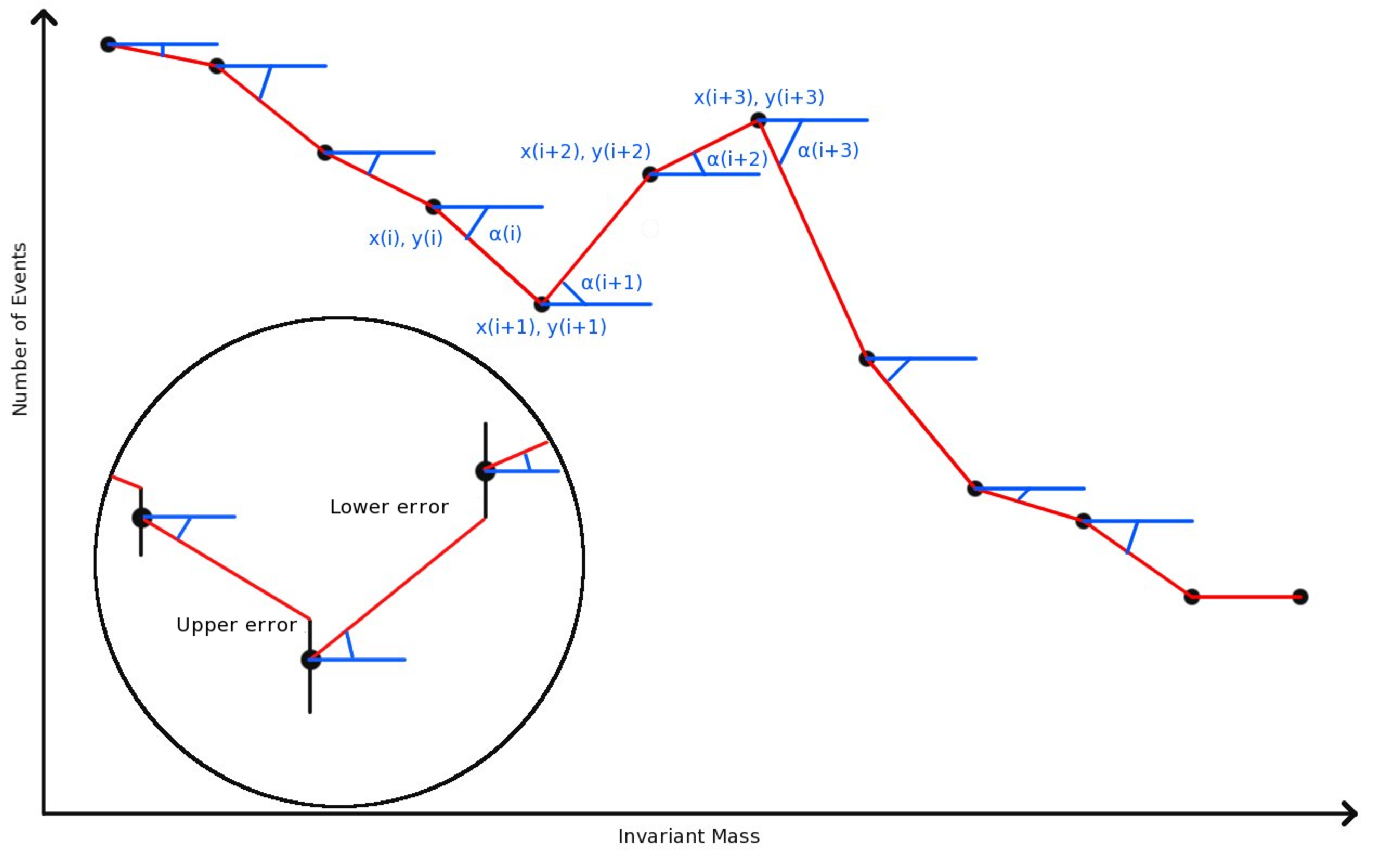,width=12cm}}
\caption
{
A graphical illustration of the NPFinder algorithm.
Each data point is characterized by a coordinate $(x_i,y_i)$, with (optional) upper and lower uncertainty on the $y_i$ values. 
See Eq.~\protect{\ref{q1}} for the definition of 
the slopes $\alpha_i$.
}
\label{graph}
\end{center}
\end{figure}

NPFinder continues to walk over data points until $\delta\alpha_{N+1}$  and $\delta\alpha_{N+2}$  are both negative, which
signifies the maximum of the peak has been reached. The double condition in Eq.~\ref{q3} is used to reinforce the peak-search robustness.
When this condition is met, NPFinder exits the peak and adds an equal number of points to the right side from the peak center.
The requirement of having the same number of points implies that the peaks
are expected to be symmetric, which is the most common case. For steeply falling distributions, such as transverse-momenta spectra or dijet mass distributions,
this assumption usually means that we somewhat underestimate the peak significance.
Figure~\ref{graph} illustrates the NPFinder algorithm for a falling invariant-mass distribution. Each point of the distribution can have 
an upper and lower statistical (or systematic) uncertainty.

After detecting all  peak candidates, NPFinder iterates through the list of possible
peaks in order to form a background for each peak. This is achieved by performing a linear regression
of points between the first and last points in the peak, i.e. applying the function $y_i=m x_i + b$, where
$m$ and $b$ are the slope and intercept of the linear regression, which in this case is rather trivial as it is performed
via the two points only. It should be noted that the linear regression is also performed taking into account uncertainties:
$$
m= \frac{ (y_{2} + \delta y_{2}) - (y_{1} + \delta y_{1})}{x_{2}- x_{1}},   
$$ 
where $y_{1}$ is the first point of the peak, $y_{2}$ is the final point of the peak, and 
$\delta y_{1}$ and $\delta y_{2}$ are their
statistical uncertainties, respectively. 
Here the statistical
uncertainties are added  in order to always be on the conservative side in estimation of the background level under the peak.
The intercept parameter then is $b=y_{1} + \delta y_{1} - m x_1$.  

It should be mentioned that the technique of the peak finding considered above is somewhat 
similar to that discussed for $\gamma$-ray applications \cite{Mariscotti1967309}. 
But there are several important differences of NPFinder compared to this algorithm: 
NPFinder can detect peaks of arbitrary shapes (not only a Gaussian-shaped peaks as in \cite{Mariscotti1967309}), no 
fitting or smoothing procedure is used, and statistical and systematic 
uncertainties for data points are included during the peak-finding procedure. The algorithm \cite{Mariscotti1967309} was not tested
since its source code is not publicly available.

Finally, NPFinder uses the background points to calculate the statistical significance of each
peak in a given histogram. This is done by summing up the differences $r_i$ of the original points in a peak with
respect to the calculated background points, and then dividing this value by it's own square root.
For a given peak, it can be approximated by:
$$
\sigma= \frac{\sum r_i}{\sqrt{{\sum r_i}}},  
$$
where the sum runs over all points in the peaks. 
The algorithm runs over an input histogram or graph, builds a list of peaks and estimates their statistical significance.
A typical statistically significant peak in this approach has $\sigma>5-7$. 
A first peak is usually ignored as it corresponds to the kinematic peak of background distributions.

Below we illustrate the above approach by 
generating fully inclusive $pp$ collision events 
using the PYTHIA generator \cite{Sjostrand:2006za}. The required integrated luminosity was 200 $\mathrm{pb}^{-1}$.  
Jets are reconstructed with the anti-$k_T$ algorithm \cite{Cacciari:2008gp}
with  a distance parameter of 0.6 using the cut $p_T>100$~GeV.  
Then,  the dijet invariant-mass distribution is calculated and the NPFinder 
finder is applied using the  parameter $\Delta=1$.  As expected, no peaks with $\sigma > 5$  were  found.

\begin{figure}[htp]
\begin{center}
\mbox{\epsfig{file=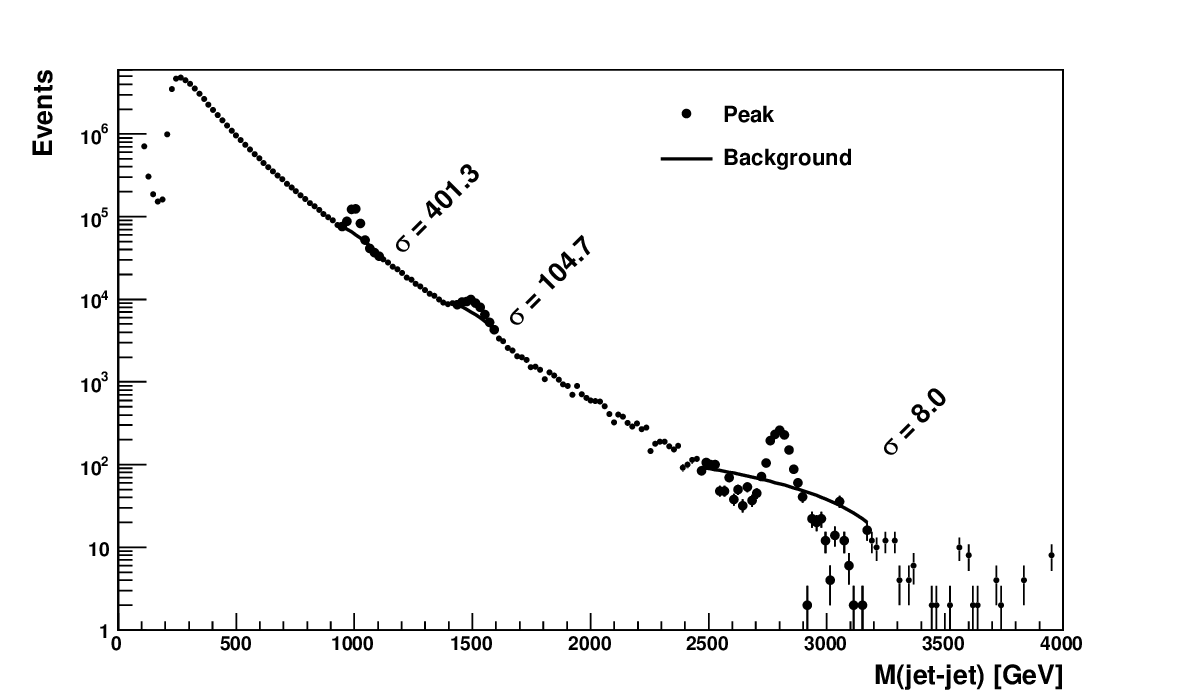,width=15cm}}
\caption
{
Invariant mass of two jets generated with the PYTHIA Monte Carlo model. 
Several peaks seen in this figure were added using Gaussian distributions with different widths
and peak values (see the text). The peaks 
are found using the NPFinder algorithms which also estimates their statistical-significance values as discussed in the text.
}
\label{jetjet}
\end{center}
\end{figure}

Next, a few fake peaks were generated using  Gaussian distributions with different peak positions and widths. The peaks were added
to the original background histogram.  
Figure~\ref{jetjet} shows an example with 3 peaks generated at 1000~GeV (20~GeV width, 200000 events), 1500~GeV (50 GeV widths, 30000 events)
and at 2800~GeV (40~GeV widths, 1200 events). The algorithm found all three peaks and gave correct estimates of their positions, widths 
and approximate statistical significance using the input parameter $\Delta=1$. 

For a comparison, the same distribution was used to test the {\sc TSpectum} package of the ROOT program discussed in the introduction.
It was found that {\sc TSpectum} can also detect such peaks, but several iterations with a visual examination of the data have been required to adjust 
the free parameters of this algorithm,  $\sigma$ (an effective sigma of searched peaks)  and the amplitude of the expected peaks.
After the first {\sc TSpectrum} pass, an additional analytic fit  was required to  determine the statistical significance of each peak.
This approach was found to be difficult to implement in a fully automatic peak search.

It should be noted that the peak statistical significance of the proposed non-parametric method might be smaller
than that calculated using more conventional approaches, such as those based on a $\chi^2$ minimisation 
with appropriate background and signal functions. This can be due to the assumption on the symmetric form of the extracted peaks, the linear approximation for 
the background under the peaks,  and
the way in which the uncertainty is incorporated in the peak-significance calculation. An influence of  experimental resolution
can also be an issue \cite{PhysRevD.76.074025} which can only be addressed via correctly identified signal and background functions.
Such drawbacks are especially well seen for the highest-mass peak shown in Figure~\ref{jetjet} where a statistical fluctuation
to the right of the peak pulls the background level up compared to the expected falling  shape. 
Given the approximate nature of the statistical significance calculations
which only serve to trigger attention of analyzers who need to study the found peaks in more detail,
the performance of the algorithm was found to be reasonable.

It should be noted that there is a correlation between the peak width and the input parameter $\Delta$: a detection of broader peaks
typically requires a  smaller value of $\Delta$ (which  can be as low as $0.2$).  

In conclusion, a peak-detection algorithm has been developed which
can be used for extraction of statistically significant peaks in
event-counting distributions taking into account statistical (and potentially systematic) uncertainties.
The method can be used for new physics searches in high-energy particle experiments where a correct treatment of
such uncertainties is one of the most important issues.  
The non-parametric peak finder has only one free parameter  which is fairly  independent of input background distributions. 
The algorithm was tested and found to perform well.
The code is implemented in the Python programming language with the graphical output using either ROOT (C++) \cite{root}
or SCaVis (Java) \cite{chekanov2010scientific}. The code example is available for download \cite{code}.

\section*{Acknowledgements}
We would like to thank J.~Proudfoot for discussion and comments. 
The submitted manuscript has been created by UChicago Argonne, LLC, 
Operator of Argonne National Laboratory ("Argonne"). 
Argonne, a U.S. Department of Energy Office of Science laboratory, 
is operated under Contract No. DE-AC02-06CH11357.

\newpage
\bibliographystyle{./Macros/l4z_pl}
\def\bibname{\Large\bf References}
\def\refname{\Large\bf References}
\pagestyle{plain}
\bibliography{biblio}

\end{document}